\documentclass[prl,twocolumn,superscriptaddress]{revtex4}
\usepackage{graphicx}
\usepackage[utf8]{inputenc}
\usepackage{amsfonts}
\usepackage{amssymb}
\usepackage{amstext}
\usepackage{amsmath}

\begin{document}
\title{Augmenting the thermal flux experiment: a mixed reality approach with the  HoloLens}
\author{M.\,P.\,Strzys}
\affiliation{Department of Physics, Physics Education Group, University of Kaiserslautern, D--67653 Kaiserslautern, Germany}
\author{S.\,Kapp}
\affiliation{Department of Physics, Physics Education Group, University of Kaiserslautern, D--67653 Kaiserslautern, Germany}
\author{M.\,Thees}
\affiliation{Department of Physics, Physics Education Group, University of Kaiserslautern, D--67653 Kaiserslautern, Germany}
\author{P.\,Lukowicz}
\affiliation{DFKI, Embedded Intelligence Laboratory, Trippstadter Str. 122, D--67663 Kaiserslautern, Germany}
\author{P.\,Knierim}
\affiliation{VIS, University of Stuttgart, Germany}
\author{A.\,Schmidt}
\affiliation{VIS, University of Stuttgart, Germany}
\author{J.\,Kuhn}
\affiliation{Department of Physics, Physics Education Group, University of Kaiserslautern, D--67653 Kaiserslautern, Germany}

%\address{Fachbereich Physik, Technische Universität Kaiserslautern, D--67653 Kaiserslautern, Germany}

\begin{abstract}
\noindent This article may be downloaded for personal use only. Any other use requires prior permission of the author and AIP Publishing. \\ The following article appeared in \textit{The Physics Teacher} 55, 376 (2017) and may be found at \\ doi: \url{http://dx.doi.org/10.1119/1.4999739}
\end{abstract}
\maketitle

In the field of Virtual Reality (VR) and Augmented Reality (AR) technologies have made huge progress during the last years \cite{Schm16,Sand15,Hock16} and also reached the field of education \cite{Sant14,Kuhn16}. The virtuality continuum, ranging from pure virtuality on one side to the real world on the other \cite{Milg94} has been successfully covered by the use of immersive technologies like head-mounted displays, which allow to embed virtual objects into the real surroundings, leading to a Mixed Reality (MR) experience. In such an environment digital and real objects do not only co-exist, but moreover are also able to interact with each other in real-time. These concepts can be used to merge human perception of reality with digitally visualized sensor data and thereby making the invisible visible. As a first example, in this paper we introduce alongside the basic idea of this column \cite{Kuhn12} an MR-experiment in thermodynamics for a laboratory course for freshman students in physics or other science and engineering subjects which uses physical data from mobile devices for analyzing and displaying physical phenomena to students. 

\section{Theoretical background}
The paradigm experiment for heat conduction in metals can be realized with a metallic rod, heated on one side while simultaneously cooled on the other \cite{Parr75}. Our setup allows to observe the heat flux through the rod, directly on the real physical object, using a false-color representation. Moreover, also additional representations as, e.g., graphs and numerical values, can be included as digital augmentations to the real experiment, allowing for a just-in-time evaluation of physical processes.

If the rod is perfectly isolated, after some equilibration time the system will reach a steady state with a hot end at temperature $T_1$, a cold end at temperature $T_2$, and a constant spatial gradient along the rod axis, which allows to calculate the thermal conductivity constant $\lambda$ of the material according to 
\begin{align}\label{lambd}
\lambda = \frac{L}{A\,(T_1-T_2)}\dot{Q},
\end{align}
if the constant heating power $\dot{Q}$ and the dimensions, i.e. length $L$ and cross-section $A$, of the rod are known. 
%In general in one dimension---we choose $x$ along the axis of the rod---the temperature distribution $T(x,t)$ obeys the heat equation
%\begin{align}\label{heateq}
%\frac{\partial}{\partial t}T(x,t)=\frac{\lambda}{\varrho c} \frac{\partial^2}{\partial x^2}T(x,t)
%\end{align}
%where $\varrho$ is the density and $c$ the specific heat capacity of the material. 
%According to Fourier's Law the heat flux $\dot{Q}$, i.e. the amount of energy that flows in $x$-direction along the rod through an area $A$ per time, is given by 
%\begin{align}\label{fourier}
%\dot{Q}=-\lambda\,A\,\frac{\partial}{\partial x}T(x,t).
%\end{align}
Here it is assumed that the full heating power $\dot{Q}$ applied to the warm end of the rod will be removed on the cold end by cooling. 

% such that according to \eqref{fourier} a linear temperature gradient is established. Thus, the thermal conductivity constant can be computed according to
%\begin{align}\label{lambd}
%\lambda = \frac{L}{A\,(T_1-T_2)}\dot{Q},
%\end{align}
%where $L$ is the lenght of the rod and $T_i$ are the two end-temperatures.
%This, however, only holds for a perfectly isolated rod. 

If the rod is not isolated, it moreover is possible to calculate the loss of heat to the environment according to $h = \alpha^2 \lambda AL$, using the decline factor $\alpha$ extracted from an exponential fit to the experimental data \cite{note1}.

%However, in the case of heat exchange with the environment, modeled as heat bath with constant temperature $T_0$, a Newtonian cooling term $-h(T-T_0)/(cm)$ \cite{Parr75}, with $h$ being the heat transfer coefficient, $m$ the mass and $c$ the specific heat capacity of the sample, has to be added to the heat equation of the system, leading to an exponential solution \cite{Parr75} $T(x,t) = T_0 + (T_1-T_0)\textrm{e}^{-\alpha x}$ for the temperature profile in this case. If the coefficient $\alpha$ is determined experimentally, the heat transfer coefficient can be calculated according to
%\begin{align}\label{heq}
%h = \alpha^2 \lambda AL.
%\end{align}
%Therefore, also the loss of heat to the environment can be determined experimentally.

\section{Experimental design}

\begin{figure}[t]
 \begin{center}
 \includegraphics[width=0.9\linewidth]{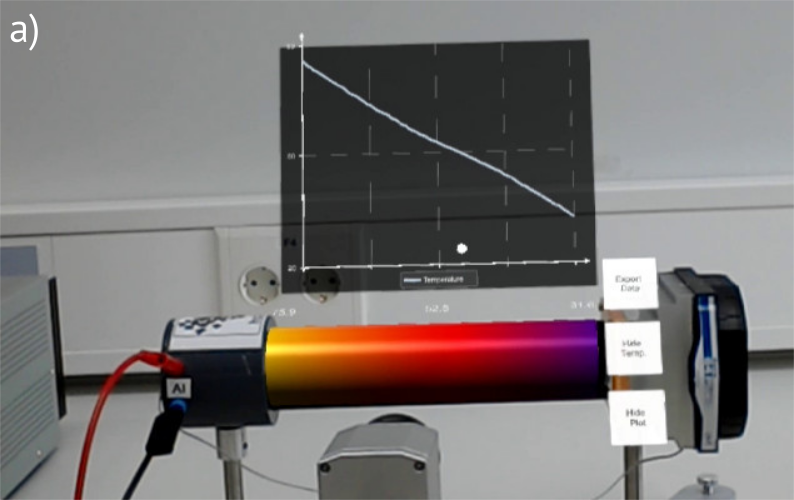}
 \includegraphics[width=0.9\linewidth]{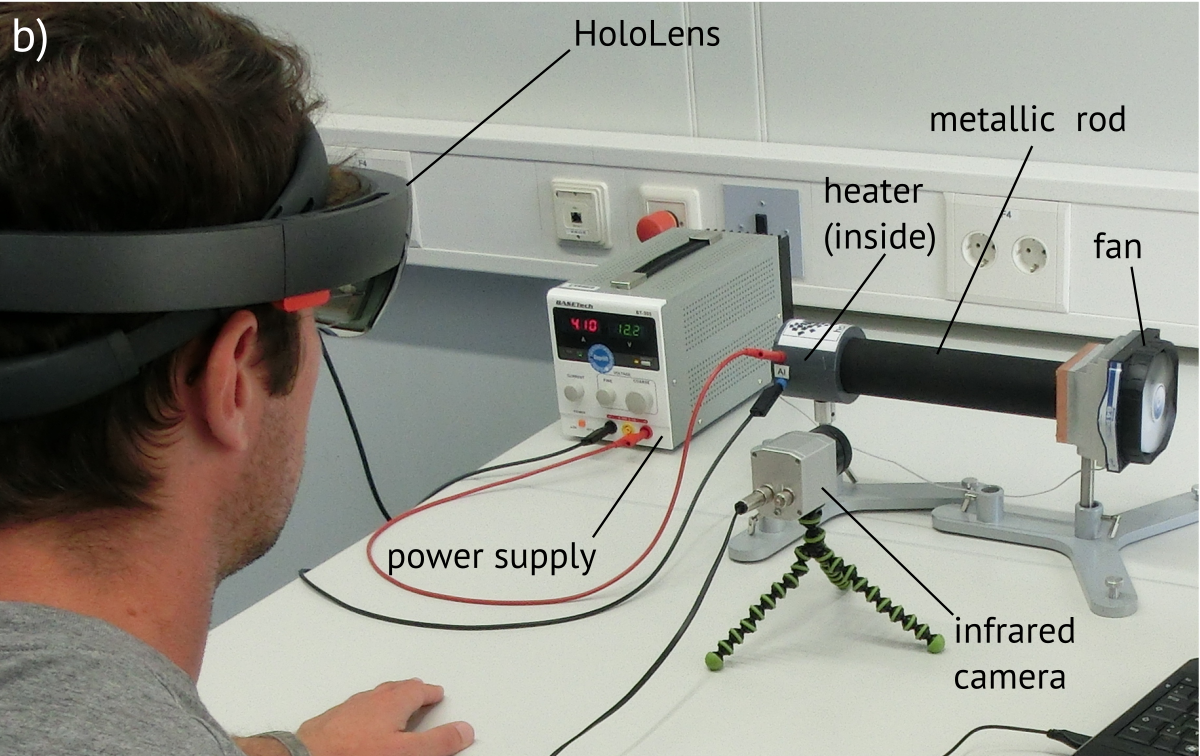}
 \includegraphics[width=0.9\linewidth]{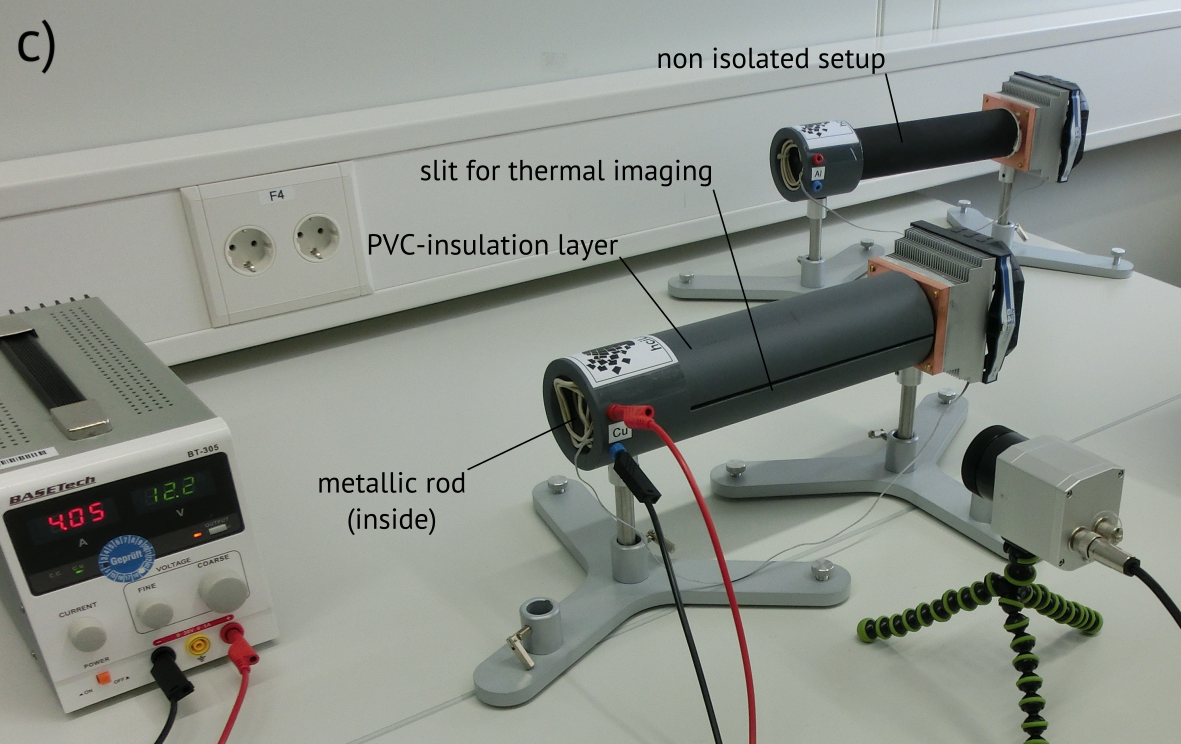}
 \end{center}
 \caption{a) Experimental setup (non isolated) with MR experience; augmented representations: false-color representation of temperature along the rod, numerical values at three points above the rod, temperature graph; b) Experimental setup (non isolated) and user wearing a HoloLens; c) isolated setup}
 \label{setup}
\end{figure}

Our heat conduction experiment consists of cylindrical metal sample rods with a length of $L=26$\,cm and a diameter of $d=5$\,cm made of Aluminum and Copper, respectively (cf. Fig.~\ref{setup}.b). The sample is heated at one end with a cartridge heater and cooled at the other by a standard CPU fan. The isolated version moreover has a PVC insulation layer with a 3\,mm slit along the rod, to allow for thermal imaging of the rod inside  (cf. Fig.~\ref{setup}.c). The temperature data finally is extracted from thermal images taken with an infrared camera placed in front of the sample. Each pixel of the image along a fixed line in axial direction yields one temperature value, such that the whole spatial distribution can be captured by a single shot. The data is then passed to the Microsoft HoloLens via WiFi, where the visualization is done. In the current state, an App provides three different representations of the data: false-color image of the temperature values projected as ``hologram'' \cite{note2}  directly onto the sample cylinder, numerical values at three pre-defined points and a temperature graph as a function of the position along the rod hovering above the setup, cf. Fig.~\ref{setup}.a. The user may switch the numerical and graph representation on and off at will; moreover, it is possible to export the data as csv-file at any time for later analysis. These functions can be executed with the help of virtual buttons (cf. the three white squares in Fig.~\ref{setup}.a) projected at the right end of the rod which can be selected by the gaze and triggered by hand gestures. 

The benefit of the MR-experiment setup is the possibility of keeping track of the real physical devices and representations of the data simultaneously and in real-time as the representations are continuously updated with new data from the camera. The false color representations allows to experience an otherwise invisible quantity, like in this case temperature, with human senses, thus extending perception to new regimes. 
%Based on multimedia learning theory, it can be assumed that time and spatial contiguity of the holographic projection directly onto the real object effectively supports the learning process and avoids the split attention effect that would occur, if other display types, like tablets or even computer screens, would be used for the virtual augmentation \cite{Maye09}. Moreover, real world annotations reduce the cognitive load of the students such that a larger fraction of the short term memory may be used during the cognitive process \cite{Sant14}. Additionally the MR setup automatically implements a contextual visualization of the data by design \cite{Sant14}. In combination with the real-time presentation of the data, we thus expect a more effective interrelation of theory and experiment, as e.g., the equilibration process can directly be followed watching the temperature graph or the false color representation and be related to theoretical predictions, which otherwise, in classical laboratory class setups, is only possible in hindsight while analyzing experimental data outside the lab. 
Furthermore, direct feedback is implemented, such that students get an immediate impression of effects of the experimental parameters.

%Since in many laboratory courses it is common for the students to work at the same experiment simultaneously in groups of two or even more, another important feature of the current design is the possibility of cooperation not only in the real, but also in the virtually augmented world. In the case of this setup this is ensured as all students attending the experiment and wearing a HoloLens are able to see and to work with the same augmentations. Therefore, they are not only able to discuss the experimental progress during the session, but also the virtual annotations and evaluations presented in their shared MR experience.

\section{Experimental Results}

%Although the main focus of this setup is on the experimental phase, a standard analysis  of the data should be carried out by the students afterwards to deepen their understanding of the connections. 
The exported data can be analyzed using standard instruments like spreadsheet programs. In the isolated case, the problem reduces to a linear fit of the temperature graph, yielding the slope and thus, with the help of the constant heating power $\dot{Q} = 50\,{\rm W}$ used in the experiments via \eqref{lambd} the thermal conductivity constant $\lambda$.
\begin{figure}[t!]
 \begin{center}
 \includegraphics[width=.5\linewidth]{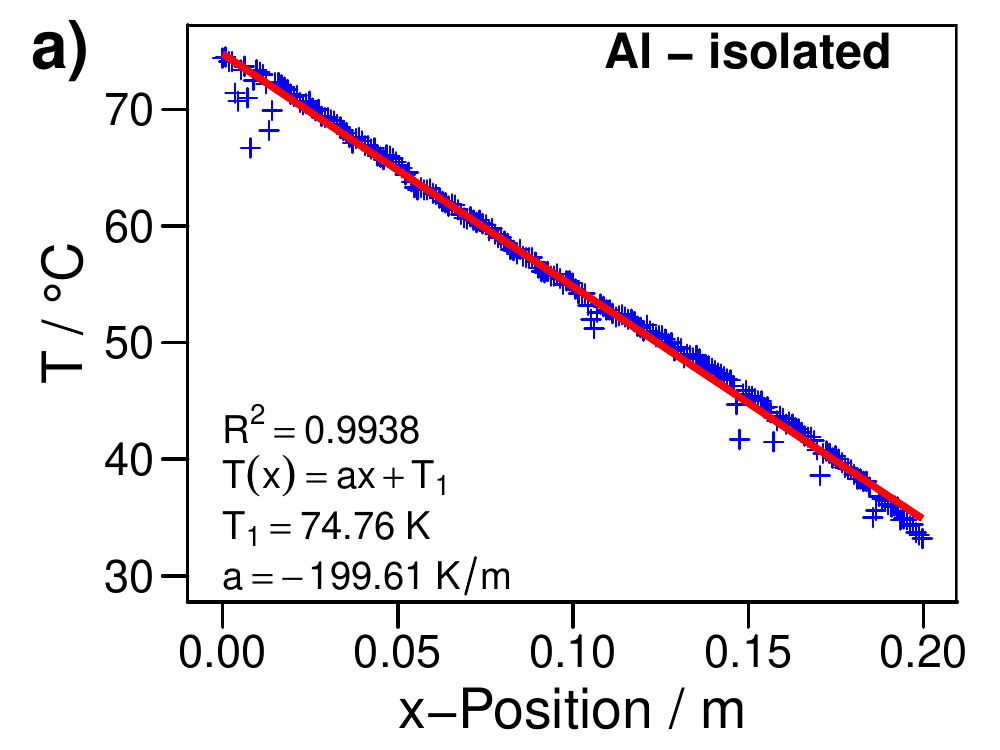}\includegraphics[width=.5\linewidth]{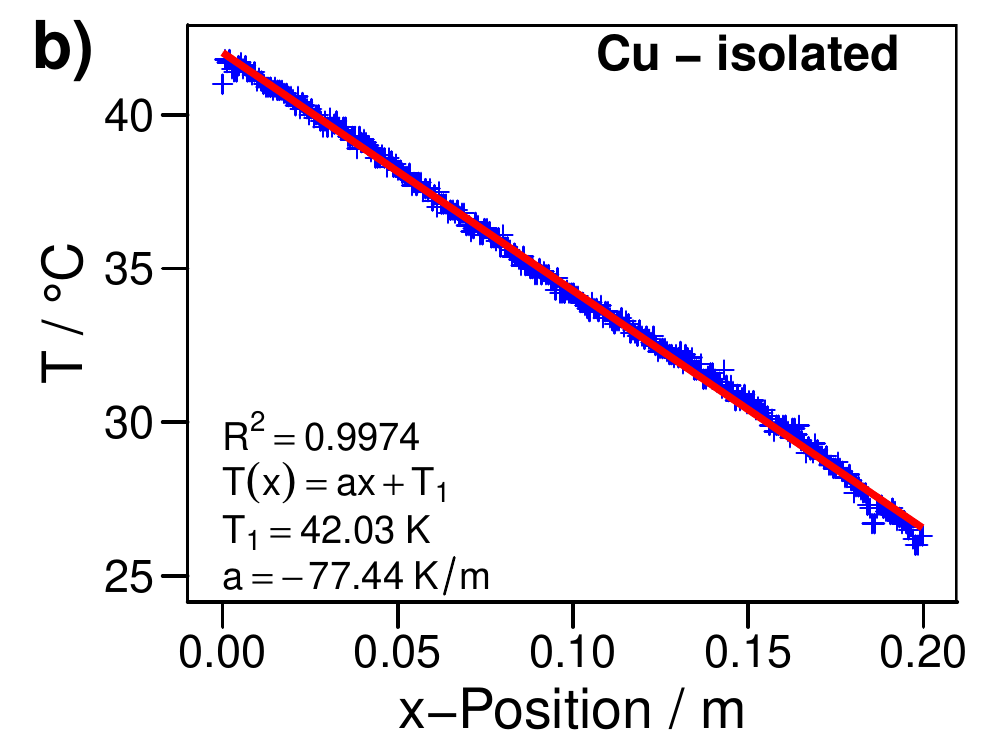}\\
 \includegraphics[width=.5\linewidth]{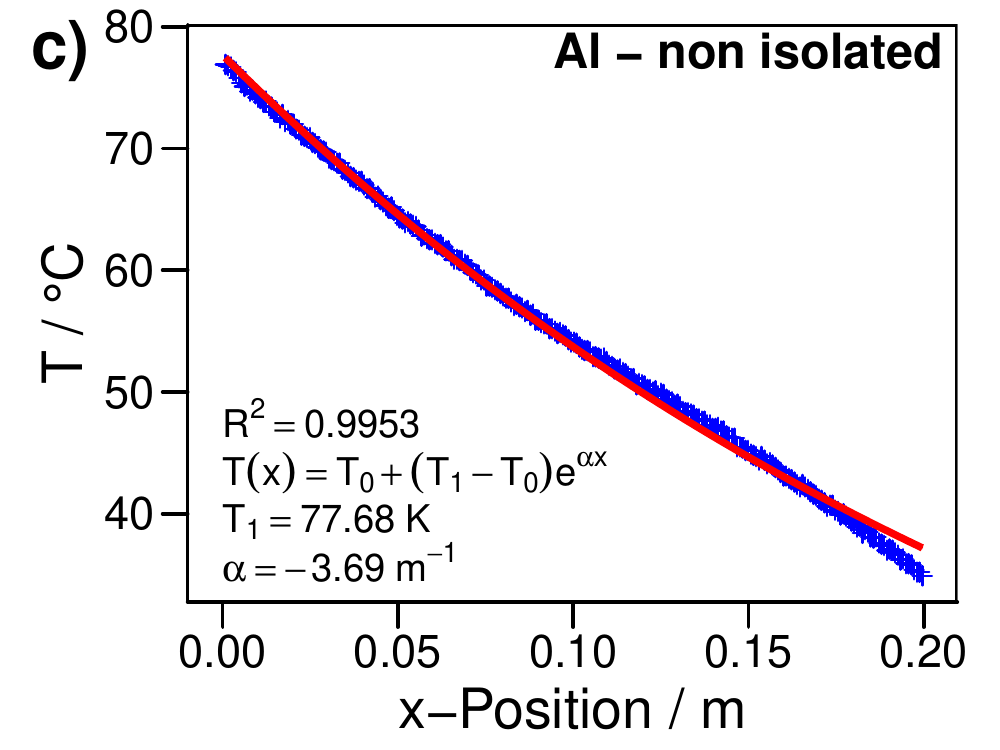}\includegraphics[width=.5\linewidth]{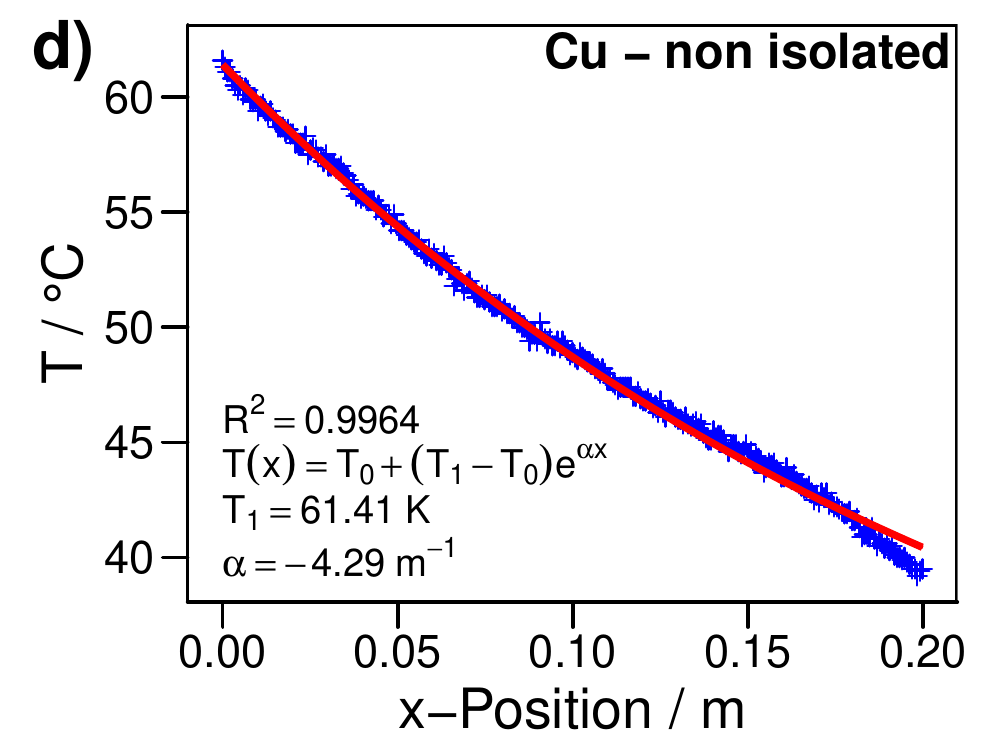}
% a)\hspace*{4cm} b)\hspace*{4cm}
 \end{center}
 \caption{Temperature graph after equilibration and linear fit for an isolated Al-rod (a), Cu-rod (b); Temperature graph after equilibration and exponential fit for a non-isolated Al-rod (c), Cu-rod (d).}
 \label{graph}
\end{figure}
In our tests (see Fig.~\ref{graph}.a,b) the isolated setup yielded a value of $\lambda_{\rm Al}=(128\pm 11)\,{\rm Wm}^{-1}{\rm K}^{-1}$ for the Aluminum rod and $\lambda_{\rm Cu}=(329\pm 27)\,{\rm Wm}^{-1}{\rm K}^{-1}$ for the Copper rod. As expected, due to the non perfect isolation of the samples, these values are smaller than the reference values found in literature, $\lambda_{\rm Al, lit}= 235\,{\rm W/(m\,K)}$ and $\lambda_{\rm Cu, lit}= 401\,{\rm W/(m\,K)}$, as the effective heat flux $\dot{Q}$ along the rod is reduced by loss to the environment.

With the help of the non-isolated rod the heat transfer coefficient $h$ can be determined using the exponential decline factor $\alpha$ (see Fig.~\ref{graph}.c,d) %according to \eqref{heq} 
yielding $h_{\rm Al} = (0.72 \pm 0.01)\,{\rm WK}^{-1}$ and $h_{\rm Cu} = (2.5 \pm 0.01)\,{\rm WK}^{-1}$, respectively. These values are also underestimated for the same reasons, however, compared to $h_{\rm Al,lit} = (1.32 \pm 0.01)\,{\rm WK}^{-1}$ and $h_{\rm Cu,lit} = (3.05 \pm 0.01)\,{\rm WK}^{-1}$,  where the literature values for $\lambda$ were used, they are reasonable regarding the very simple insulation of the setup.

\section{Conclusion}

The MR experimental setup for a thermal flux experiment presented here, sheds new light on an experiment well known in physics laboratory courses. The main focus of this design is to visualize the invisible and thus to extend human perception to new regimes, e.g., temperature and heat, thereby strengthening the connection between theory and experiment. In this realization the MR setting not only has the advantage of intrinsic contextuality, but also spatial and time contiguity which is supposed to support the learning process of the students \cite{Craw01,Maye10}. Moreover, the just-in-time evaluation of the data yields the possibility for the students to directly examine the process itself and the parameter involved, and immediately compare the outcome to theoretical predictions which we believe to enhance the links between theory and experiment. Under that perspective the effort to achieve possibly more exact numerical values for quantities like the thermal conductivity therefore seem to be less important in this setting. Instead, the technical support during the experimental phase will give students the possibility to thoroughly examine the relationship between cause and effect and thus deepen their physical understanding.

Support from the German Federal Ministry of Education and Research (BMBF) via the project ``Be-greifen'' \cite{begreifen} is gratefully acknowledged.

\bibliographystyle{unsrt}

%\bibliography{ar}

\begin{thebibliography}{1}
\bibitem{note2}
The HoloLens does not actually use interference-based holograms but rather projections to the transparent head-mounted displays of the device. However, the term hologram is used by the manufacturer to describe these.
\end{thebibliography}

\end{document}